\documentclass[twocolumn,showpacs,preprintnumbers,amsmath,amssymb]{revtex4}

\usepackage{graphicx}
\usepackage{dcolumn}
\usepackage{bm}

\begin{document}

\title{Gas-like state of $\alpha$ clusters around $^{16}$O core in $^{24}$Mg}
\author{T. Ichikawa$^1$, N. Itagaki$^1$, T. Kawabata$^2$, Tz. Kokalova$^3$, and W. von Oertzen$^{4}$
}

\affiliation{
$^1$Yukawa Institute for Theoretical Physics, Kyoto University,
Kitashirakawa Oiwake-Cho, 606-8502 Kyoto, Japan} 

\affiliation{
$^2$Division of Physics and Astronomy, Kyoto University,
Kitashirakawa Oiwake-Cho, 606-8502 Kyoto, Japan} 

\affiliation{
$^3$School of Physics and Astronomy, University
of Birmingham, Edgbaston, B15 2TT, Birmingham, UK
}

\affiliation{
$^4$Helmholtz-Zentrum Berlin, Glienicker Str. 100, D-14109 Berlin, Germany}

\date{\today}

\begin{abstract}
We have studied gas-like states of $\alpha$ clusters around 
an $^{16}$O core in $^{24}$Mg 
based on a microscopic $\alpha$-cluster model.
This study was performed by
introducing a Monte Carlo technique 
for the description of the THSR
(Tohsaki Horiuchi Schuck R\"{o}pke)
wave function, and the coupling effect to other
low-lying cluster states was taken into account.
A large isoscalar monopole ($E0$) transition strength
from the ground to the gas-like state is discussed.
The gas-like state of two  $\alpha$ clusters
in $^{24}$Mg 
around the $^{16}$O core appears slightly below
the 2$\alpha$-threshold energy.


\end{abstract}

\maketitle

\section{Introduction}
In the last couple of years, gas-like states comprised of 
$\alpha$ clusters in atomic nuclei have attracted increased 
interest \cite{Tohsaki,Schuck,Funaki,Matsumura,Yamada, Kurokawa}.
It is basically considered that all of the $\alpha$-clusters in
such states
occupy the same $0s$-orbital, which has a spatially extended 
distribution \cite{Schuck}.
This situation is well expressed by introducing the so-called THSR 
(Tohsaki Horiuchi Schuck R\"{o}pke) wave function, 
where the oscillator parameter is large,
which is completely different from the normal $0s$-orbital for
each nucleon. The most plausible candidates for such gas-like states 
of $\alpha$ clusters are the second $0^+$ state of 
$^{12}$C$^*$ (3$\alpha$) at $E_x = 7.65$ MeV and more generally,  
states around the $N \alpha$-threshold energies.
The squared overlap between the wave function of a microscopic 
cluster model and the THSR wave function 
has been found to be
more than 90$\%$ for $^{12}$C \cite{Funaki,Matsumura}, 
which suggests that the single THSR wave function is 
a good approximation for the description of the gas-like state. 
Furthermore, a candidate for the gas-like state of 4$\alpha$ clusters 
around the threshold energy in $^{16}$O has been studied
by both theoretical and experimental approaches \cite{Tohsaki,Funaki,Wakasa}.
 
Lately, the research into gas-like states comprised of $\alpha$ clusters
moved onto the second stage.
For instance, the possibility of gas-like states of $\alpha$
clusters around a core in heavier nuclei has been 
suggested \cite{Kokalova-1,Kokalova-2,vOe,Ogloblin}.
The coherent emission of $\alpha$ clusters
from the compound nucleus 
in heavier nuclei has been reported \cite{Kokalova-1}, 
which leads us to a hypothesis 
that gas-like states of $\alpha$ clusters can be formed 
not only in $^{12}$C and $^{16}$O
but also in heavier nuclei with some core. 
Using the $^{28}$Si+$^{24}$Mg reaction, the compound states of $^{52}$Fe 
have been populated, 
and the $^8$Be($0^+_1$) and $^{12}$C($0^+_2$) emissions
from these states have been observed, 
which are much enhanced compared to the sequential $\alpha$-emission. 
From a statistical model point of view, 
it is natural to consider that the emitted second $0^+$ state 
of $^{12}$C is formed inside the Coulomb 
barrier of the compound nucleus. 
The enhancement of the emission of 
the gas-like states of $\alpha$ clusters could be due to 
the lowering of the effective 
Coulomb barrier for the condensed states \cite{Kokalova-2},
since the kinetic energy of the emitted $^{12}$C
in coincidence with $\gamma$-emission has been 
observed to be much smaller than the energy sum of three $\alpha$'s in
the sequential 3$\alpha$ emission. 

To study the possibility of gas-like states of $\alpha$ clusters 
in heavier nuclei from the theoretical side,
we have introduced 
a Monte Carlo technique for the description of the THSR wave function, 
which is called the ``virtual THSR" wave function \cite{VT}.
We have shown the possibility of a three $\alpha$ state
around a $^{40}$Ca core \cite{40Ca}.
The three $\alpha$ cluster state around $^{40}$Ca 
has an energy below the Coulomb barrier top energy
with a spatial extension 
comparable to the second $0^+$ state of $^{12}$C.

Recently, it has been proposed that the strong enhancement 
of isoscalar monopole (E0) transitions can be a measure
of the existence of cluster structure \cite{Kawabata}.  
For instance, the presence of cluster states in $^{13}$C
has been suggested by measuring the isoscalar E0 transitions
from the ground $1/2^-$ state induced by the $^{13}$C$(\alpha, \alpha')^{13}$C
reaction \cite{Sasamoto}. The obtained cross-sections are much larger than those
of the shell-model calculations. The result suggests that protons
and neutrons are coherently excited and  that they have spatially
extended distribution in the excited states.
From the theoretical side, 
the relation between the monopole transition strength and the cluster
structure has also been discussed \cite{YIO, Yoshida-2,Uegaki, HO, Yamada-1}. 
The basic idea arises from the Bayman-Bohr theorem \cite{BB59},
which shows that the lowest representation of the shell-model
contains a component of the lowest $SU(3)$ representation of the cluster states
and the monopole operator connects the lowest $SU(3)$ state
and the well-developed cluster states.

The experimental search for the large monopole strength
to gas-like cluster state(s) of $\alpha$ clusters around an $^{16}$O core
is on going, and we analyze such structure from the theoretical point of view.
In this paper we study the gas-like state of two $\alpha$ clusters around an $^{16}$O
core in $^{24}$Mg, and here, the coupling effect between the gas-like $\alpha$ cluster
state and the normal cluster states in the low-lying region is taken into account.

\section{Method}
The original THSR 
wave function for the $\alpha$-condensed state has the following form:
\begin{eqnarray}
\Phi = && \int d \vec{R_1} d\vec{R_2} \cdot \cdot \cdot d\vec{R_n} \nonumber \\
&& {\cal A} \ \ G(\vec r_1, \vec R_1)G(\vec r_2, \vec R_2)G(\vec r_3, \vec R_3) \cdot \cdot \cdot G(\vec r_n, \vec{R_n}) \nonumber \\
&& \times \exp[ -(\vec R_1^2 + \vec R_2^2 + \vec R_3^2 \cdot \cdot \cdot \vec R_n^2 )/\sigma^2 ] \nonumber \\
= && {\cal A} \ \ \prod_{i=1}^n \int d \vec R_i \ G(\vec r_i, \vec R_i) \exp[-\vec R_i^2 / \sigma^2],
\end{eqnarray}
where $\cal A$, $G$($\vec r_i$, $\vec R_i$) = $\exp[-\nu(\vec r_i - \vec R_i)^2]$,
and $\sigma$ 
are the antisymmetrizer, the wave function for the $i$-th 
$\alpha$-cluster centered at $\vec R_i$, 
and the oscillator parameter of the $\alpha$-condensation,
respectively. 
The four nucleons (proton spin-up, proton spin-down, neutron spin-up, and neutron spin-down) 
in the $i$-th $\alpha$-cluster share the common Gaussian center parameter $\vec R_i$.

To simplify this wave function, 
we introduce the ``virtual THSR" wave function 
in the following way \cite{VT}:
\begin{equation}
\Psi^\sigma = \sum_{k = 1}^m P^\pi P^J_{MK} \Psi_k^\sigma,
\end{equation}
\begin{equation}
\Psi_k^\sigma = [ {\cal A} \ G(\vec r_1, \vec R_1)G(\vec r_2, \vec R_2)G(\vec r_3, \vec R_3) \cdot \cdot \cdot G(\vec r_n, \vec{R_n}) ]_k.
\end{equation}
Here, the integral over the Gaussian center parameters $\{ \vec R_i \}$ 
in the original THSR wave function (in Eq. (1))
is replaced by the sum of many Slater determinants. 
The Gaussian center parameters $(\{ \vec R_i \})$ are randomly generated
by the weight function $W$ with a Gaussian shape:
\begin{equation}
W(\vec R_i) \ \propto \ \exp[- \vec R_i^2 / \sigma^2]. 
\end{equation}
With increasing ensemble number, the distribution of $\{ \vec R_i \}$
approaches a Gaussian with $\sigma$ width parameter.  
Thus, it can be considered 
that the integration in the original THSR wave function (see Eq. (1)) 
is performed by using a Monte Carlo technique 
for the virtual THSR wave function, 
and the wave function agrees with the original THSR wave function
when the number of Slater determinants ($m$ in Eq. (2)) increases. 

In the present case, we add a $^{16}$O core
and discuss the two  $\alpha$ state around it.
Thus, each basis state (Eq. (3)) becomes
\begin{equation}
\Psi_k^\sigma = [ {\cal A} \ \phi(^{16}{\rm O}) 
\ G(\vec r_1, \vec R_1)G(\vec r_2, \vec R_2)]_k.
\end{equation}
Here the $^{16}$O core $(\phi(^{16}{\rm O}))$ is described by
a tetrahedron configuration of four $\alpha$-clusters with small
relative distances, and only the Gaussian center parameters
of two $\alpha$ clusters around the $^{16}$O 
core $(\vec R_1, \vec R_2)$ 
are randomly generated with the weight function $W(\vec R_i)$. 
The center of mass of each basis state $\Psi_k$ is shifted 
to the origin after generating the wave functions of $\alpha$ clusters
around $^{16}$O.
The projection onto good 
parity ($P^\pi$) and angular momentum ($P^J_{MK}$)
is performed numerically. Here, $\pi$ is positive parity and $J = M = K = 0$.
The number of mesh points for the Euler angle integral is $16^3 = 4096$.
The Gaussian size of each nucleon ($\nu$ in $G$($\vec r_i$, $\vec R_i$) = $\exp[-\nu(\vec r_i - \vec R_i)^2]$) 
is taken to be $\nu = 1/2b^2$, $b=$ 1.46 fm.

In the present study,
we calculate the coupling effect between
the gas-like cluster states described by the virtual THSR wave function
and normal cluster states. The normal cluster states $\{ \Psi^j_{nc} \}$
are described by many different configurations of $^{16}$O+$\alpha$+$\alpha$ for $^{24}$Mg.
The total wave function $\Psi$ is therefore
\begin{equation}
\Psi =  \sum_j c^j_{nc} \Psi^j_{nc} + \sum_\sigma c^\sigma \Psi^\sigma,
\end{equation}
and the coefficients $\{ c^j_{nc}  \}$ and $\{ c^\sigma \}$ are determined
by diagonalizing the Hamiltonian matrix.

The Hamiltonian operator $(\hat{H})$ has the following form:
\begin{equation}
\hat{H}=\sum_{i=1}^{A}\hat{t}_i-\hat{T}_{c.m.}+\sum_{i>j}^{A}\hat{v}_{ij},
\end{equation}
where $\hat{t}_i$ is the kinetic energy of $i$-th nucleon, and 
the center-of-mass kinetic energy ($\hat{T}_{c.m.}$) is exactly removed.
Here, the two-body interaction $(\hat{v}_{ij})$ 
includes the central part and the Coulomb part. 
We use the following Volkov No.2 effective $N-N$ potential \cite{Vol}:
\begin{equation}
V(r) = (W - MP^\sigma P^\tau) \sum_{k=1,2} V_k \exp(-r^2/c_k^2),
\end{equation}
where $W = 1-M$ ($M$: Majorana exchange parameter). 
It is known that although $M$ $\sim$ 0.6 reproduces the
$\alpha$-$\alpha$ scattering phase shift, larger $M$ values
are needed for the structure calculation
beyond $^{12}$C, and here $M$ is chosen to be 0.63
to yield a reasonable binding energy of $^{24}$Mg.
Using the present interaction, the lowest state of $^{16}$O+2$\alpha$ ($^{24}$Mg)
is calculated to be $-16.24$ MeV from the three-body threshold energy,
when we perform a GCM (generator coordinate method) calculation by 
superposing the basis states and diagonalizing
the Hamiltonian, compared with the experimental value of $-14.046$ MeV.

\section{Results}

We start the discussion with the virtual THSR wave function
for the gas-like state of $\alpha$ clusters around $^{16}$O ($\Psi^\sigma$).
The energy convergence of $^{16}$O+$\alpha$+$\alpha$ ($^{24}$Mg)
as a function of number of basis states ($m$ in Eq. (2)) is shown in Fig. 1.
The energy is measured from the $^{16}$O+$\alpha$+$\alpha$ threshold.
Our previous application of the virtual THSR wave function 
for the three $\alpha$ ($^{12}$C) case has shown that
the wave function with $\sigma$ = 4 fm is found to give
a reasonable root mean square radius for the second $0^+$ of $^{12}$C
compared with that obtained by other approaches \cite{VT}.
In the present case, the result of $\sigma$ = 4 fm 
(also that of $\sigma$ = 5 fm)
shows the converged energy below the threshold due to the presence of
the $^{16}$O core.
In the case of $\sigma = 3$ fm, the converged energy
is much lower ; $\sim -6$ MeV below the threshold.

\begin{figure}
\includegraphics[width=\linewidth]{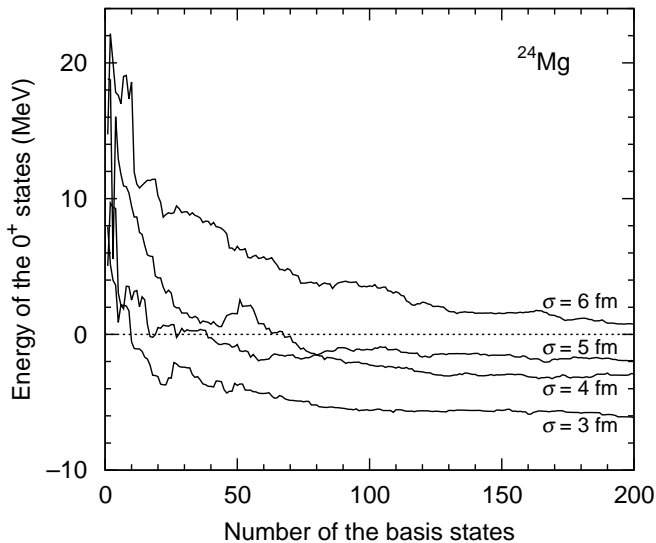}
\caption{
The energy convergence of two $\alpha$ clusters  
around the $^{16}$O core ($^{24}$Mg) for the case of $\sigma$ = 3, 4, 5 and 6 fm.
The energy is measured from the $^{16}$O+$2\alpha$ threshold.
}
\end{figure}

Next we calculate the coupling effect between
the gas-like cluster states described by the virtual THSR wave function 
($\Psi^\sigma$, $\sigma$ = 3, 4, 5 and 6 fm) 
and normal cluster states ($\Psi^j_{nc}$).
The left column of Fig. 2 shows the $0^+$ levels of $^{24}$Mg
obtained by diagonalizing the Hamiltonian consisting of 200 different 
configurations of $^{16}$O+$\alpha$+$\alpha$. 
The energy convergence of $\{ \Psi^j_{nc} \}$ is shown in Fig. 3.
In addition to this normal cluster states of $\{ \Psi^j_{nc} \}$,
we superpose virtual THSR wave functions with the $\sigma$ values
of $\sigma$ = 3, 4, 5 and 6 fm, and the result obtained after the coupling
is shown as the right column of Fig. 2. 
In principle, the model space of $\Psi_{nc}$ covers that of $\Psi^\sigma$
if we prepare an infinite number of different configurations,
however the energies of the calculated  7th $0^+$ state 
at $-4.267$ MeV is  influenced
by the coupling with the virtual THSR wave function.
This state is considered to have a large component of gas-like 
$\alpha$ clusters. Because of the presence of the $^{16}$O core,
the 7th $0^+$ state appears below the threshold energy.

\begin{figure}
\includegraphics[width=\linewidth]{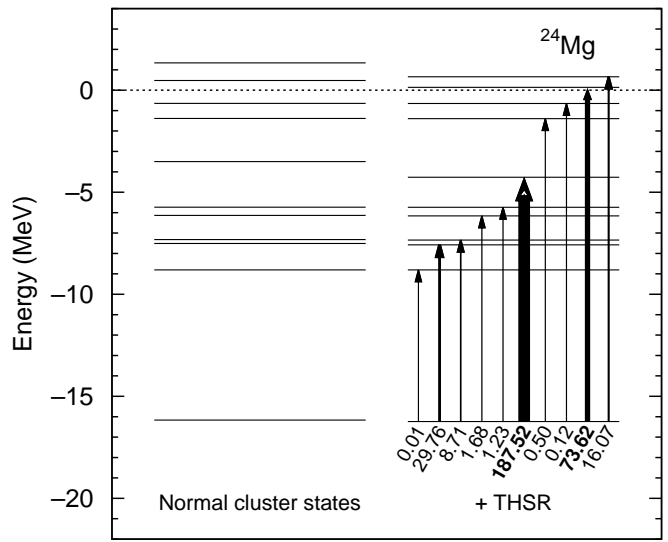}
\caption{
The eleven lowest $0^+$ states of $^{24}$Mg obtained by the $^{16}$O+$\alpha$+$\alpha$ model.
The left column shows the levels 
obtained by diagonalizing the Hamiltonian consisting of 200 different 
configurations of normal cluster states ($\Psi^j_{nc}$).
The right column shows the results after 
 superposing the virtual THSR wave functions ($\Psi$) with the $\sigma$ values
of $\sigma$ = 3, 4, 5 and 6 fm in addition to $\Psi^j_{nc}$.
The arrows and numbers (fm$^4$) 
show the E0 transition strength
from the ground state.
}
\end{figure}

\begin{figure}
\includegraphics[width=\linewidth]{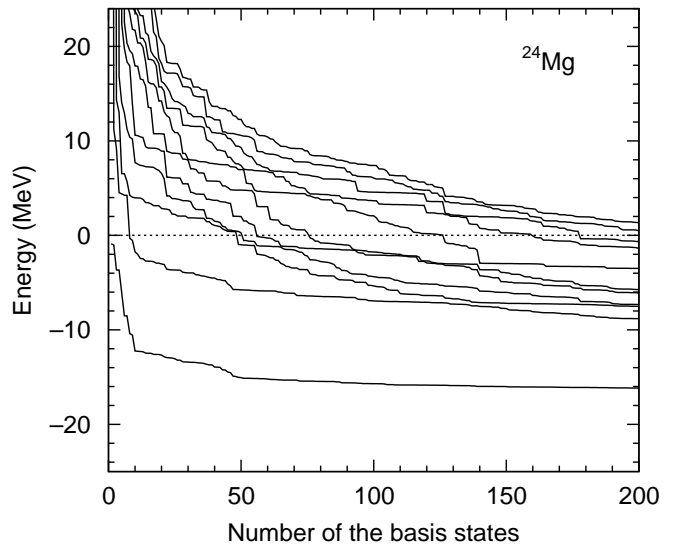}
\caption{
The energy convergence of 
eleven lowest $0^+$ states of $^{24}$Mg obtained by 
the basis states of normal cluster states 
with the $^{16}$O+$\alpha$+$\alpha$ configurations ($\Psi^j_{nc}$).
}
\end{figure}

\begin{table}
\begin{center}
\caption{
The energies of the eleven lowest $0^+$ states of $^{24}$Mg 
measured from the $^{16}$O+$\alpha$+$\alpha$ threshold (MeV) and
squared isoscalar monopole (E0) transition strength from the ground state (fm$^4$).
}
\begin{tabular}{||c|c|}
\hline
\hline
energy (MeV) & squared monopole strength (fm$^4$) \\
\hline
$-16.24$ & ---  \\
$-8.807$ & 0.009  \\
$-7.580$ & 29.76  \\
$-7.344$ & 8.709  \\
$-6.157$ & 1.684  \\
$-5.739$ & 1.231  \\
$-4.267$ & 187.5  \\
$-1.393$ & 0.503  \\
$-0.650$ & 0.118  \\
 0.140   & 73.62  \\
 0.658   & 16.07  \\
\hline
\end{tabular}
\end{center}
\end{table}

The energies of the eleven lowest $0^+$ states  of $^{24}$Mg
measured from the $^{16}$O+$\alpha$+$\alpha$ threshold (MeV) and
squared  isoscalar monopole (E0) transition strength from the ground state (fm$^4$) are
 summarized in Table I.
The calculated monopole transitions from the ground state to the 7th $0^+$ state at $-4.267$ MeV 
shows strong enhancement (187.5 fm$^4$).
The result shows that the state which has large components of gas-like cluster basis states
is strongly excited by the E0 transition,
which could be observed in experiments as a signature of well developed $\alpha$ cluster state.


\begin{table}
\begin{center}
\caption{
The energies of the eleven lowest $0^+$ states of $^{24}$Mg 
measured from the $^{16}$O+$\alpha$+$\alpha$ threshold (MeV) and
squared overlap with the THSR wave functions for the two $\alpha$
clusters around $^{16}$O ($\sigma$ = 3,4,5 and 6 fm).
}
\begin{tabular}{||c|c|c|c|c|}
\hline
\hline
energy (MeV) & $\sigma$ = 3 fm  & $\sigma$ = 4 fm   &  $\sigma$ = 5 fm & $\sigma$ = 6 fm \\
\hline
$-16.24$ & 0.234  & 0.114  & 0.094   &  0.046 \\
$-8.807$ & 0.000  & 0.000  & 0.000   &  0.000 \\
$-7.580$ & 0.113  & 0.070  & 0.058   &  0.025 \\
$-7.344$ & 0.023  & 0.014  & 0.012   &  0.005 \\
$-6.157$ & 0.007  & 0.006  & 0.005   &  0.003 \\
$-5.739$ & 0.003  & 0.002  & 0.002   &  0.001 \\
$-4.267$ & 0.429  & 0.426  & 0.392   &  0.263 \\
$-1.393$ & 0.001  & 0.002  & 0.002   &  0.002 \\
$-0.650$ & 0.000  & 0.000  & 0.000   &  0.000 \\
 0.140   & 0.038  & 0.092  & 0.083   &  0.087 \\
 0.658   & 0.016  & 0.047  & 0.046   & 0.005  \\
\hline
\end{tabular}
\end{center}
\end{table}

The squared overlaps between each of the $0^+$ states and
the THSR wave functions for the two $\alpha$
clusters around $^{16}$O ($\sigma$ = 3, 4, 5, and 6 fm)
are shown in Table II. The ground state has a large
overlap with $\sigma = 3$ fm, because the 
limit of $\sigma$ = 0 fm corresponds to the limit
of the Elliot SU(3) state. It is confirmed that 
the 7th $0^+$ state has a considerable amount
of overlap with the THSR wave functions with larger
$\sigma$ values (e.g. about 40\% with $\sigma$ = 5 fm).
The 10th $0^+$ state at 0.140 MeV
just around the $^{16}$O+$\alpha$+$\alpha$
threshold energy, which also has a large E0 transition strength 
from the ground state (76.32 fm$^4$ in Table I),
has a certain overlap with the THSR wave functions with
large $\sigma$ values.

\section{Conclusion}

We have studied $\alpha$ cluster
states around an $^{16}$O core, which has been hinted at
experimentally. The approach is based on a
microscopic $\alpha$ cluster model. This was performed
by
introducing a Monte Carlo technique 
for the description of the THSR
(Tohsaki Horiuchi Schuck R\"{o}pke)
wave function, 
which is called the ``virtual THSR" wave function.

The two $\alpha$ cluster state around an $^{16}$O 
core in $^{24}$Mg 
has an energy below the threshold energy,
when the spatial extension is
comparable to that of the second $0^+$ state of $^{12}$C.
The character of this state survives after imposing the coupling 
with other cluster states in lower energy region. The 7th $0^+$ state
below the $^{16}$O+$\alpha$+$\alpha$ threshold 
has significant overlaps with the THSR wave functions
for large $\sigma$ values.
The calculated isoscalar monopole (E0) transition from the ground state
to this state shows strong enhancement, which could be observed
as an experimental signature of a well developed $\alpha$
gas-like state.

\begin{acknowledgments}
The authors would like to thank C. Wheldon and P. Schuck for important
suggestions. Numerical computation in this work was carried out at the 
Yukawa Institute Computer Facility using the new SR16000 system.
This work was done in the Yukawa International Project for Quark-Hadron Sciences (YIPQS).
\end{acknowledgments}


\begin{references}

\bibitem{Tohsaki}
A. Tohsaki, H. Horiuchi, P. Schuck, and G. R\"{o}pke, 
Phys. Rev. Lett. {\bf 87}, 192501 (2001).

\bibitem{Schuck}
P. Schuck, Y. Funaki, H. Horiuchi, G. R\"{o}pke, A. Tohsaki and T. Yamada,
Nucl. Phys. {\bf A738}, 94 (2004).

\bibitem{Funaki} 
Y. Funaki, A. Tohsaki, H. Horiuchi, P. Schuck, and G. R\"{o}pke, 
Phys. Rev. C {\bf 67}, 051306(R) (2003).\\
Y. Funaki, A. Tohsaki, H. Horiuchi, P. Schuck, and G. R\"{o}pke, 
Mod. Phys. Lett. A {\bf 21}, 2331 (2006).\\
Y. Funaki $et \ al.$,
Phys. Rev. C {\bf 80}, 064326 (2009).

\bibitem{Matsumura}
H. Matsumura and Y. Suzuki, Nucl. Phys. {\bf A739}, 238 (2004).

\bibitem{Yamada} 
T. Yamada and P. Schuck,
Phys. Rev. C {\bf 69}, 024309 (2004).

\bibitem{Kurokawa} 
C. Kurokawa and K. Kat\=o,
Phys. Rev. C {\bf 71}, 021301(R) (2005).

\bibitem{Wakasa} 
T. Wakasa $et \ al.$,   
Phys. Lett. B {\bf 653}, 173 (2007).



\bibitem{Kokalova-1} 
Tz. Kokalova $et \ al.$, Eur. Phys. J. A {\bf 23}, 19 (2005).

\bibitem{Kokalova-2} 
Tz. Kokalova, N. Itagaki, W. von Oertzen, and C. Wheldon, 
Phys. Rev. Lett. {\bf 96}, 192502 (2006).

\bibitem{vOe}
W. von Oertzen, Eur. Phys. J. A {\bf 29}, 133 (2006). 

\bibitem{Ogloblin}
A.A. Ogloblin, S.A. Goncharov, T.L. Belyaeva, and A.S. Demyanova,
Phys. of Atom. Nucl. {\bf 69}, 1149 (2006).

\bibitem{VT}
N. Itagaki, M. Kimura, C. Kurokawa, M. Ito and W. von Oertzen,
Phys. Rev. C {\bf 75}, 037303 (2007).\\
N. Itagaki, Tz. Kokalova, M. Ito, M. Kimura, and W. von Oertzen, 
Phys. Rev. C {\bf 77}, 037301 (2008).\\
N. Itagaki, M. Ito, K. Arai, S. Aoyama and Tz. Kokalova 
Phys. Rev. C {\bf 78}, 017306 (2008). \\
S. Aoyama and N. Itagaki 
Phys. Rev. C {\bf 80}, 021304(R) (2009). 

\bibitem{40Ca}
N. Itagaki, Tz. Kokalova, and W. von Oertzen,
Phys. Rev. C {\bf 82}, 014312 (2010).

\bibitem{Kawabata}
T. Kawabata $et\ al.$,
Phys. Lett. B {\bf 646}, 6 (2007).

\bibitem{Sasamoto}
Y. Sasamoto $et\ al.$, Mod. Phys. Lett. A {\bf 21}, 2393 (2006).

\bibitem{YIO}
T. Yoshida, N. Itagaki and T. Otsuka, Phys. Rev. C {\bf 79}, 034308 (2009).

\bibitem{Yoshida-2}
T. Yoshida, N. Itagaki, and K. Kat\=o,
Phys. Rev. C {\bf 83} 024301 (2011).

\bibitem{Uegaki}
E. Uegaki $et\ al.$, Prog. Theor. Phys. {\bf 62}, 1621 (1979).

\bibitem{HO}
H. Horiuchi, Prog. Theor. Phys. Suppl. {\bf 62}, 90 (1977).

\bibitem{Yamada-1}
T. Yamada $et\ al.$, Prog. Theor. Phys. {\bf 120}, 6 (2008).

\bibitem{BB59}
B. F. Bayman and A. Bohr, Nucl. Phys. {\bf 9}, 596 (1959).

\bibitem{Vol}
A.B. Volkov, Nucl. Phys. {\bf 74}, 33 (1965).



\end{references}
\end{document}